\documentclass[final,1p,times]{elsarticle} 

\usepackage{graphicx}
\usepackage{amssymb} 
\usepackage{amsthm} 
\usepackage{lineno}
\usepackage{lineno}
\usepackage{wrapfig}

\journal{Nuclear Physics A} 

\begin{document}

\begin{frontmatter} 

\title{PHENIX Measurements of Higher-order Flow Harmonics for Identified Charged Hadrons in Au+Au collisions at $\sqrt{s_{NN}}=39-200$ GeV}

\author{Yi Gu (for the PHENIX\fnref{col1} Collaboration)}
\fntext[col1] {A list of members of the PHENIX Collaboration and acknowledgements can be found at the end of this issue.}
\address{Department of Chemistry, State University of New York, Stony Brook, NY 11794, U.S.A}


\begin{abstract} 
    The azimuthal anisotropy coefficients $v_{2,3,4}$, characterizing collective flow in Au+Au collisions, 
are presented for identified particle species as a function of transverse momentum ($p_T$), centrality ($\rm{cent}$) and 
beam collision energy ($\sqrt{s_{NN}}=0.039-0.20$ TeV). The $v_n$ values for each particle species, show little, if any, 
change over the measured beam 
energy range, and  ${v_n}/{(n_q)^{n/2}}$ vs. $KE_T/n_q$ scales to a single curve (constituent quark number ($n_q)$ scaling) for 
each $n$, over a broad range of transverse kinetic energies ($KE_T$). A comparison of $v_2(p_T)$ for individual particle species obtained 
in Au+Au collisions at $\sqrt{s_{NN}}=0.20$ TeV (RHIC) and Pb+Pb collisions at $\sqrt{s_{NN}}=2.76$ TeV (LHC), 
indicate stronger collective flow at the LHC. These flow measurements and their scaling patterns, 
can provide important additional constraints for extraction of the specific viscosity $\eta/s$. 

\end{abstract} 

\end{frontmatter} 


\section{Introduction}

Collective flow continues to play a central role in ongoing efforts to characterize the transport 
properties of the Quark Gluon Plasma (QGP) produced in heavy ion collisions at RHIC. 
An experimental manifestation of this flow is the anisotropic emission of particles in the 
plane transverse to the beam direction. This anisotropy can be characterized by the Fourier coefficients $v_n$
determined relative to the estimated participant event planes $\Psi_{n}$: 
\begin{equation}
v_{n}=\frac{\langle cos n\left(\phi- \Psi_{n}\right)\rangle}{Res\{\Psi_{n}\}} 
\label{eq:1}
\end{equation}
where $\phi$ is the azimuthal angle of an emitted particle, $Res\{\Psi_{n}\}$
is a resolution factor which accounts for the dispersion of the azimuthal angle of $\Psi_{n}$, and brackets 
denote averaging over particles and events. The anisotropic emission of particles can also be characterized equivalently
by the pair-wise distribution in the azimuthal angle difference 
($\Delta\phi =\phi_a - \phi_b$) between particle pairs with transverse 
momenta $p^a_{T}$ and $p^b_{T}$ (respectively) 
\begin{equation}
\frac{dN^{\rm{pairs}}}{d\Delta\phi} \propto \left( 1 + \sum_{n=1}2v^a_nv^b_n\cos(n\Delta\phi) \right),
\label{eq:2}
\end{equation}
\begin{wrapfigure}{r}{0.55\textwidth}
 \vspace{-13pt}
   \begin{center}
    \includegraphics[width=0.55\textwidth]{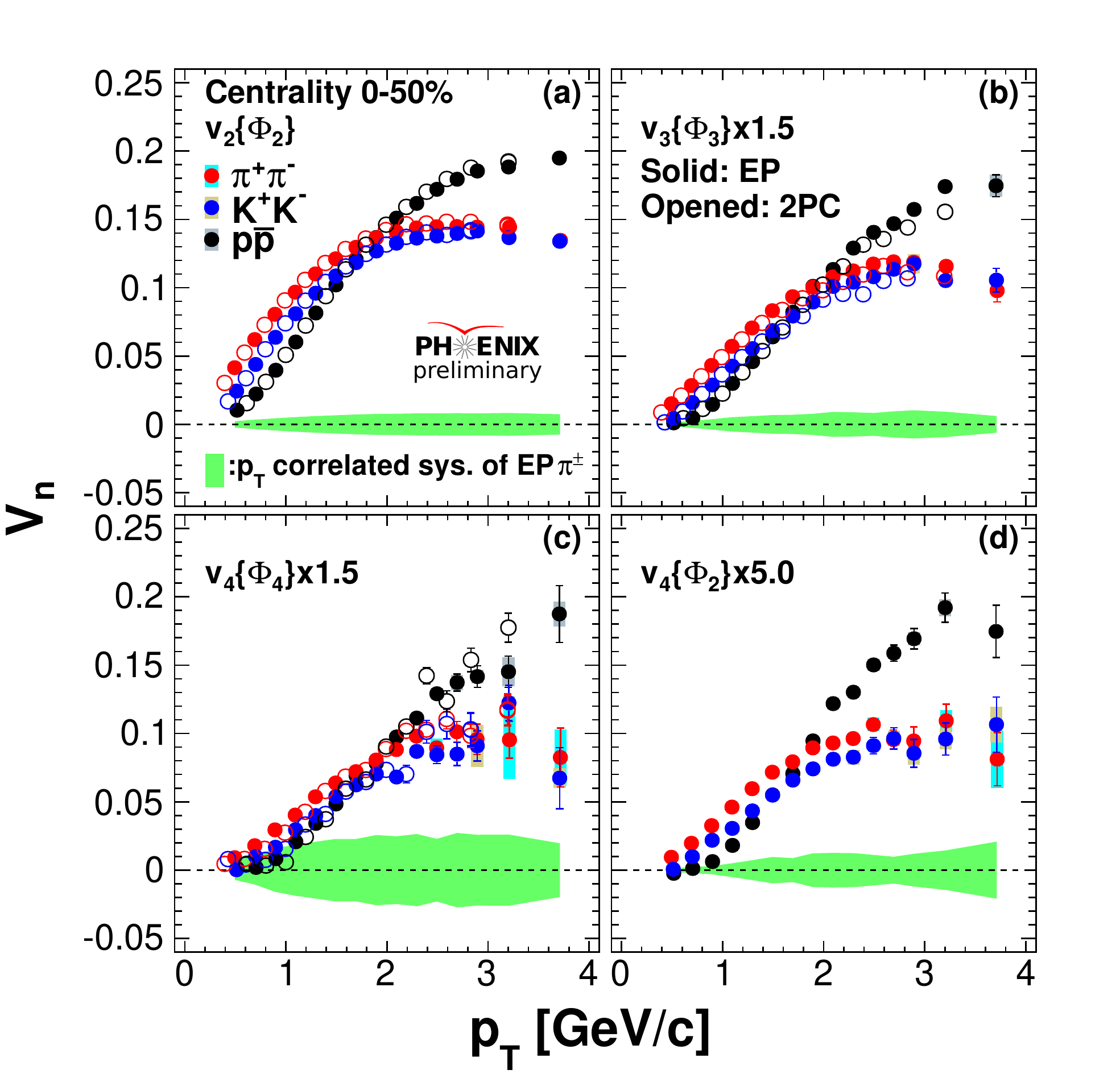}
    \end{center}
    \vspace{-20pt}
    \caption{$v_{2,3,4}$ vs. $p_T$ for $\pi^{\pm}$, $K^{\pm}$ and $\bar{p}p$ for 0-50\% central 
  Au+Au collision at $\sqrt{s_{NN}}=200$ GeV; results for the EP and 2PC methods are compared.}
  \label{fig:figure1}
\end{wrapfigure}

Significant attention has been given to the use of $v_2(\rm{cent}, p_T)$ measurements for 
the extraction of the specific shear viscosity $\eta/s$, (ie. the ratio of viscosity to 
entropy density) ~\cite{Song2011hk}. However, the uncertainty for $\eta/s$ remains large, primarily 
because of an uncertainty in the initial state geometry used in model calculations.
Recent developments suggest that the higher order flow harmonics $v_{n \ge 3}$ for inclusive charged 
hadrons, provide tighter constraints for disentangling the respective role of initial state 
geometry and  $\eta/s$ ~\cite{Lacey2011ug,PPG132}. Consequently, systematic 
measurements of the flow coefficients $v_{n}$ for identified particle species might be expected to 
provide additional constraints.

\section{Methods and Analysis}

In PHENIX, flow coefficients are extracted via the event plane method (Eq.~\ref{eq:1}) and the long-range two particle 
correlation method (2PC) (Eq.~\ref{eq:2}). The results presented at QM(and in this proceeding) are for particles produced at mid-rapidity, and reconstructed in the PHENIX central arms.

\begin{wrapfigure}{r}{0.59\textwidth}
 \vspace{-20pt}
   \begin{center}
    \includegraphics[width=0.59\textwidth]{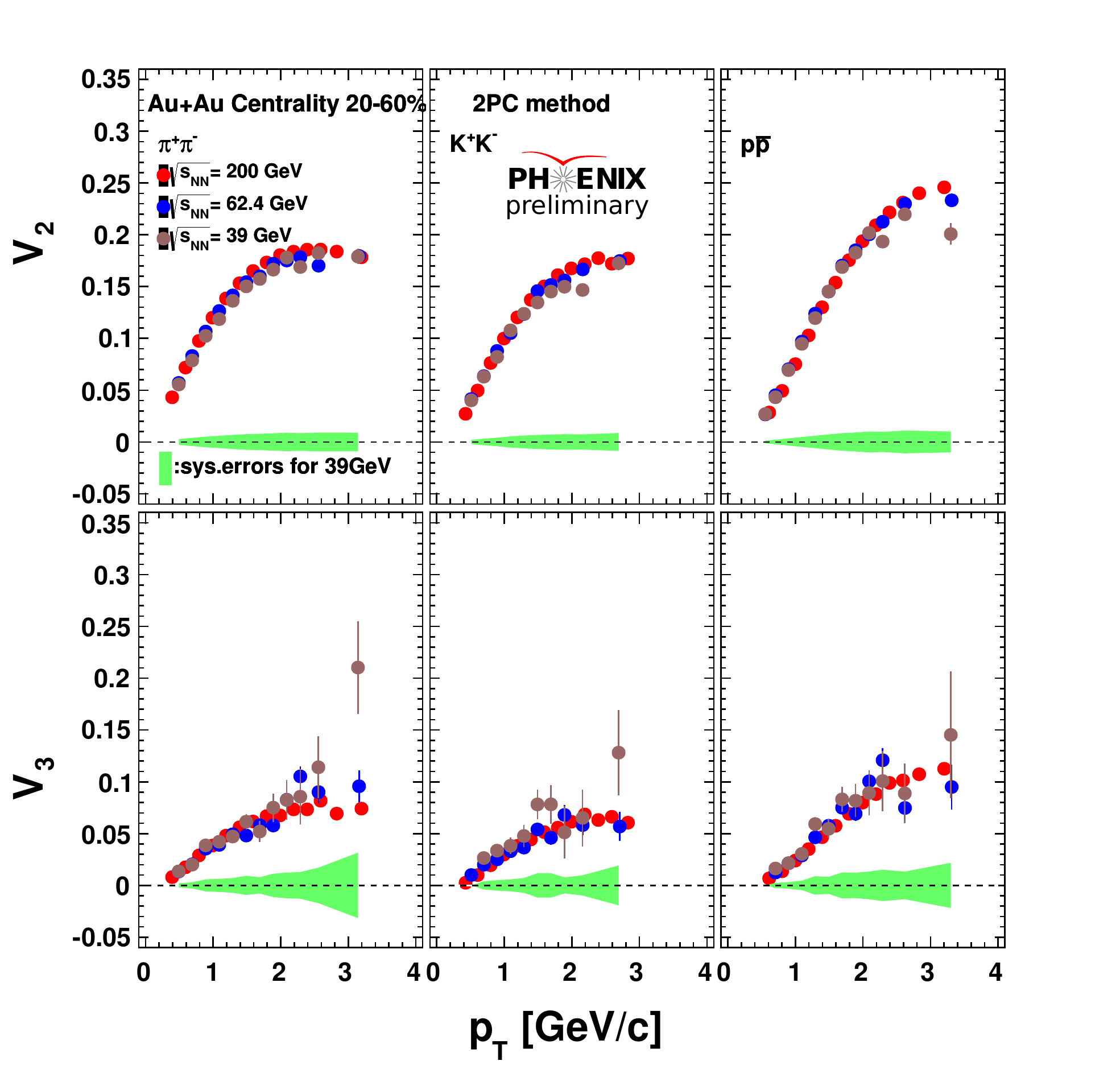}
    \end{center}
    \vspace{-20pt}
    \caption{Comparison of $v_{n}$ vs. $p_T$ for for $\pi^{\pm}$, $K^{\pm}$ and $\bar{p}p$ respectively, 
  at $\sqrt{s_{NN}}=39, 62.4$ and 200 GeV. Results are shown for the 20-60\% centrality cut.}
  \label{fig:figure2}
\end{wrapfigure}
For the event plane method (EP) (cf. Eq.\ref{eq:1}), several event plane detectors, with  
different pseudo-rapidity ($\eta_p$) coverages were employed: 
the Raction Plane detector(RxnP: $| \eta_p| =1.0-2.8$), 
the Muon Piston Calorimeter(MPC: $| \eta_p| =3.1-3.7$) and the Beam-Beam Counter(BBC: $|\eta_p| =3.1-3.9$). These event 
planes enabled robust consistency checks, as well as reliable systematic error estimates.
The 2PC method (cf. Eq.\ref{eq:2}) was performed by correlating tracks in the central PHENIX 
arm ($|\eta_p| \le 0.35$) with hits in RxnP ($|\eta_p| =1.0-2.8$) to produce correlation 
functions (the ratio of the distributions for correlated pairs and uncorrelated pairs from mixed 
events). The flow coefficients were extracted by Fourier analyzing these correlation functions.


The flow coefficients $v_{2,3,4}(p_T)$ obtained with both methods, for pions ($\pi^{\pm}$), kaons ($K^{\pm}$) and (anti)protons ($\bar{p}p$) in 0-50\% central Au+Au collisions at $\sqrt{s_{NN}}=200$ GeV, are compared in Fig.~\ref{fig:figure1}. The open symbols (2PC method) and filled symbols (EP method) indicate good consistency, suggesting that ``non-flow'' correlation
(primarily from jet fragmentation and resonance decays) might be small. 
The relatively large values of $\Delta\eta_p$ between the PHENIX central arms and the event plane detectors serve to 
reduce such correlations. 
Similarly good agreement were obtained for finer (10\%) centrality cuts.

\begin{wrapfigure}{r}{0.6\textwidth}
 \vspace{-5pt}
   \begin{center}
    \includegraphics[width=0.6\textwidth]{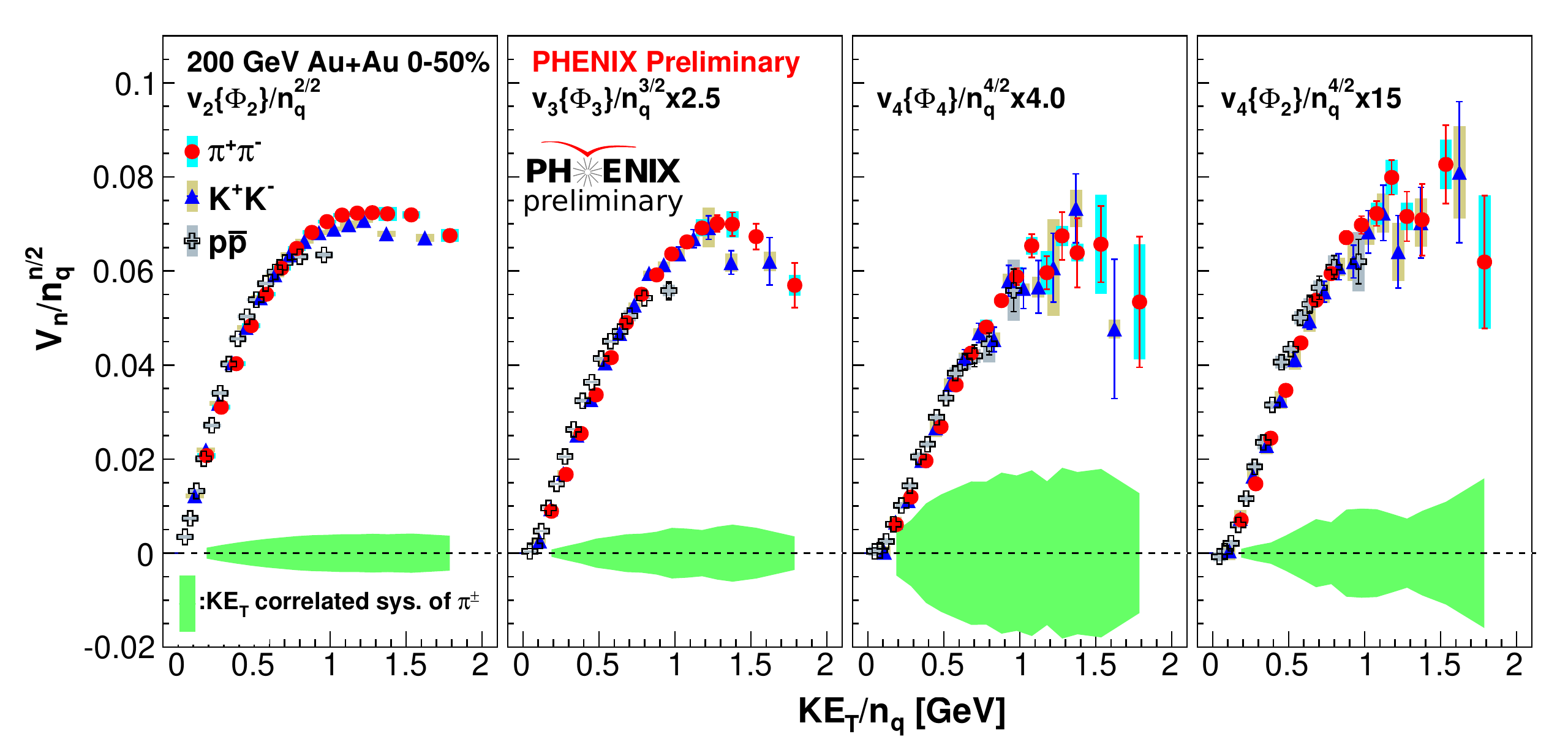}
    \end{center}
    \vspace{-20pt}
    \caption{${v_n}/{n_q^{n/2}}$vs.$KE_T/n_q$ for n=2,3,4 for 0-50\% central Au+Au collisions at 
  $\sqrt{s_{NN}}=200$ GeV via EP method}
  \vspace{-10pt}
  \label{fig:figure3}
\end{wrapfigure}
Figure~\ref{fig:figure2} compares the respective $v_{2,3}(p_T)$ values obtained at $\sqrt{s_{NN}}=39, 62.4$ 
and 200 GeV, for each particle species. Within systematic errors, the flow coefficients for 
$\pi^{\pm}$, $K^{\pm}$ and $\bar{p}p$ respectively, indicate very little, if any, change 
as the beam energy is increased. This points to a possible ``saturation'' of 
collective flow for the beam energy range $\sqrt{s_{NN}}=39-200$ GeV. Here, it is important to emphasize
that the saturation reflected in the comparison for each particle species {\em cannot} arise from the 
interplay between radial and elliptic flow which could result in a cancellation between the $v_n$ values 
for light and heavier particles to give a constant $v_n$ with beam collision energy \cite{Shen2012vn}.
Such a saturation could however, result from a softening of the equation of state.

\begin{wrapfigure}{r}{0.59\textwidth}
 \vspace{-5pt}
   \begin{center}
    \includegraphics[width=0.59\textwidth]{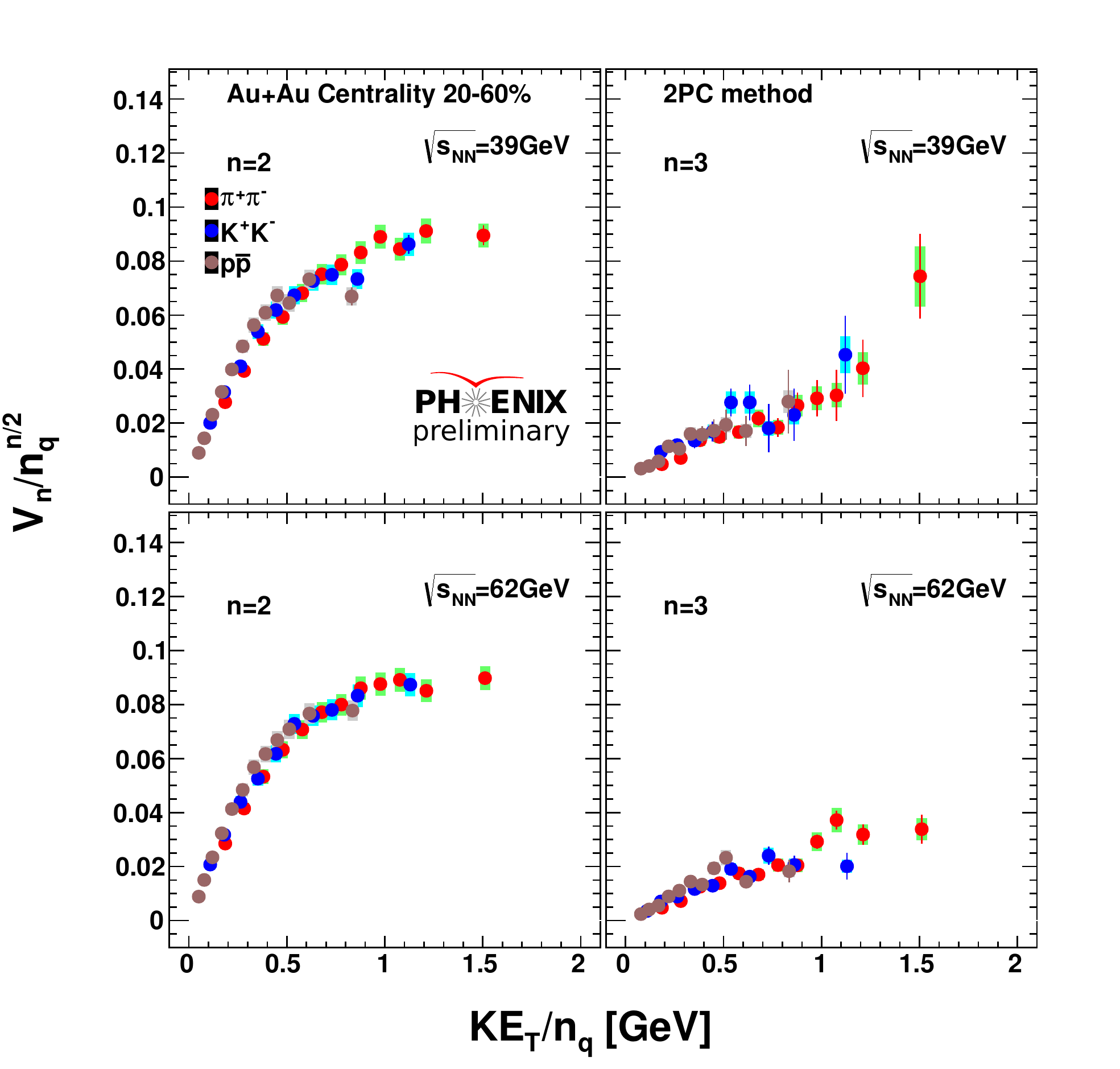}
    \end{center}
    \vspace{-20pt}
    \caption{${v_n}/{n_q^{n/2}}$vs.$KE_T/n_q$ for n=2,3 for 20-60\% central Au+Au collisions at 
  39, 62.4 GeV via the 2PC method.}
  \label{fig:figure4}
\end{wrapfigure}
The number of constituent quark (NCQ) scaling properties of these flow measurements are shown in Fig.~\ref{fig:figure3} 
and Fig.~\ref{fig:figure4} for $\sqrt{s_{NN}}=200$ GeV and 39, 62.4 GeV respectively. 
They indicate that 
${v_n}/{n_q^{n/2}}$ vs. $KE_T/n_q$ (for n=2,3,4) scale to a single curve, confirming that NCQ scaling also holds 
for the lower beam energies of $\sqrt{s_{NN}}=39$ and 62.4 GeV.
%
%

\begin{wrapfigure}{r}{0.6\textwidth}
 \vspace{-15pt}
   \begin{center}
    \includegraphics[width=0.6\textwidth]{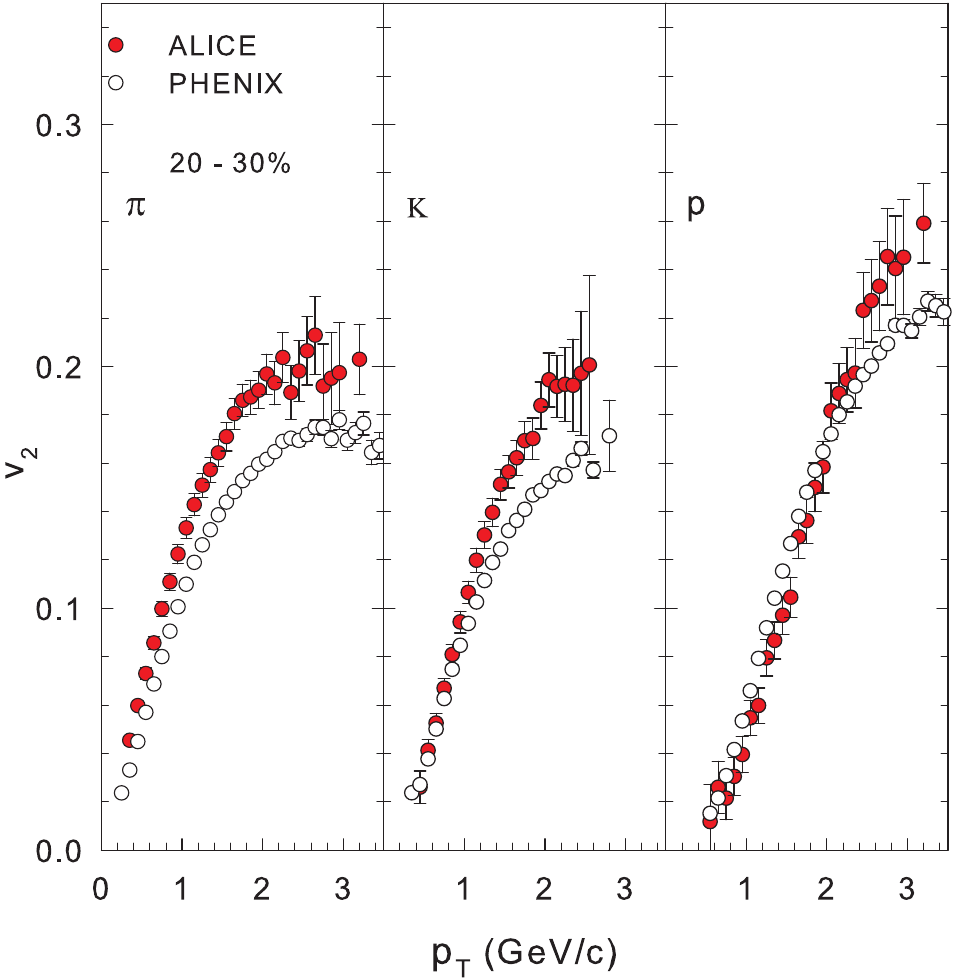}
    \end{center}
    \vspace{-10pt}
    \caption{Comparison of $v_{2}(p_T)$ measurements for Au+Au collisions at $\sqrt{s_{NN}}=0.20$ TeV RHIC (PHENIX) 
    and Pb+Pb collisions at $\sqrt{s_{NN}}=2.76$ TeV LHC (ALICE) \cite{Snellings_qm2011}. Results 
    are shown for $\pi^{\pm}$, $K^{\pm}$ and $\bar{p}p$ for the 20-30\% most central 
    collisions.}
  \label{fig:figure7}
\end{wrapfigure}

A comparison between RHIC and LHC $v_2(p_T)$ is shown for the 20-30\% most central collisions in Fig.\ref{fig:figure7}. 
It indicates a larger flow for pions and kaons at all $p_T$'s as might be expected from the significant 
energy density increase from RHIC to LHC. For (anti)protons, the LHC values are larger than the RHIC values 
for $p_T$ $\gtrsim$ 2.5 GeV/c, however this trend is inverted for lower $p_T$. The latter inversion can be attributed 
to a much larger radial flow [at the LHC] which gives a larger blueshift to the $v_2(p_T)$ values 
for (anti)protons \cite{Heinz:2011kt,Lacey:2012ma}.
Note that, in contrast to our measurements for identified particle species at RHIC, this 
interplay between radial and elliptic flow results in a subtle cancellation between increasing contributions 
from light, and decreasing contributions from heavier particles to make inclusive charged hadron $v_2$ similar 
at RHIC and the LHC. 
The reported $v_n$ measurements should provide important additional constraints 
for $\eta/s$ extraction via future model comparisons.

\section{Summary}
PHENIX has performed a new and comprehensive set of $v_n$ measurements for identified particle species at $\sqrt{s_{NN}}=39, 62.4$ and 200 GeV. The new measurements show a ``saturation'' of flow, consistent with our previous inclusive charged hadron flow measurements. However, a comparison of $v_2(p_T)$ for individual particle species obtained 
in Au+Au collisions at RHIC and Pb+Pb collisions at the LHC, indicate stronger collective flow at the LHC.
At the LHC, the interplay between radial and elliptic flow leads to a subtle cancellation between the $v_n$ values 
for light and heavier particles, to make the magnitude of inclusive charged hadron $v_n$ similar to those at RHIC.
The RHIC $v_n$ measurements show that quark number scaling (${v_n}/{(n_q)^{n/2}}$ vs. $KE_T/n_q$ for 
different particle species scale to a single curve) holds for each harmonic, for a broad range of 
transverse kinetic energies.


\section*{References}

\begin{thebibliography}{00} 
%
\bibitem{Song2011hk} H. Song, S. A. Bass, U. Heinz, T. Hirano, C. Shen, Phys.Rev. C83 (2011) 054910.
\bibitem{Lacey2011ug} R. A. Lacey, A. Taranenko, N. Ajitanand, J. Alexander, arXiv 1105.3782 (2011).
\bibitem{PPG132} Adare, A. and et al. (PHENIX Collaboration) Phys. Rev. Lett. 107 (2011) 252301.
\bibitem{Shen2012vn} C. Shen, U. Heinz, Phys.Rev. C85 (2012) 054902.
\bibitem{Heinz:2011kt} U. Heinz, C. Shen, H. Song, arXiv 1108.5323 (2012)
\bibitem{Lacey:2012ma} R. Lacey, Y. Gu, X. Gong, D. Reynolds, N. N. Ajitanand, J. M. Alexander, A. Mwai and A. Taranenko, 
        arXiv 1207.1886 (2012)
\bibitem{Snellings_qm2011} R. Snellings, J.Phys. G38 (2011) 124013.
%
\end{thebibliography}
\end{document}